\documentclass[twocolumn,showpacs,aps,prd,floatfix]{revtex4}

\usepackage{graphicx}
\usepackage{dcolumn}
\usepackage{amsmath}
\usepackage{epsfig}

\input pubboard/babarsym

\def\chicj {\ensuremath{\chi_{cJ}}\xspace}
\def\jpsiee {\ensuremath{{\jpsi\to\epem}}}
\def\jpsimm {\ensuremath{{\jpsi\to\mumu}}}
\def\pstar {\ensuremath{{p^*}}}
\def\coshel {\ensuremath{{\cos\theta_H}}}

\newcommand{\BABARPubYear}    {02}
\newcommand{\BABARPubNumber}  {04}

\newcommand{\SLACPubNumber} {9327}
\newcommand{\LANLNumber} {0207097}

\def\figurebox#1#2#3{%
    \def\arg{#3}%
    \ifx\arg\empty
    {\hfill\vbox{\hsize#2\hrule\hbox to #2{\vrule\hfill\vbox to #1{\hsize#2\vfill}\vrule}\hrule}\hfill}%
    \else
    {\hfill\epsfbox{#3}\hfill}%
    \fi}

\begin{document}

\begin{flushleft}
\babar-PUB-\BABARPubYear/\BABARPubNumber\\
SLAC-PUB-\SLACPubNumber\\
hep-ex/\LANLNumber
\end{flushleft}

\title{
Study of Inclusive Production of Charmonium Mesons in $B$ Decays
}

%
\author{B.~Aubert}
\author{D.~Boutigny}
\author{J.-M.~Gaillard}
\author{A.~Hicheur}
\author{Y.~Karyotakis}
\author{J.~P.~Lees}
\author{P.~Robbe}
\author{V.~Tisserand}
\author{A.~Zghiche}
\affiliation{Laboratoire de Physique des Particules, F-74941 Annecy-le-Vieux, France }
\author{A.~Palano}
\author{A.~Pompili}
\affiliation{Universit\`a di Bari, Dipartimento di Fisica and INFN, I-70126 Bari, Italy }
% ...
\collaboration{The \babar\ Collaboration}
\noaffiliation

\date{\today}

\begin{abstract}
The inclusive production of charmonium mesons in $B$ meson decay has
been studied in a 20.3~\invfb\ data set collected by the \babar\
experiment operating at the \FourS\ resonance.  Branching
fractions have been measured for the inclusive production of the
charmonium mesons 
\jpsi, \psitwos, \chicone, and \chictwo.  The branching fractions are
also presented as a function of the center-of-mass momentum of the mesons
and of the helicity of the \jpsi.
\end{abstract}


\maketitle

\section{Introduction}

Studies of the inclusive production of charmonium mesons in $B$ decays
provide insight into the physics of the underlying production
mechanisms.  Non-relativistic QCD (NRQCD) \cite{ref:bodwin}, which may provide 
an explanation \cite{ref:psi2sbrat} for the
unexpectedly large production of \jpsi\ and \psitwos\ mesons observed 
in $p \overline p$ collisions \cite{ref:abe},
is an example of such a mechanism.  NRQCD calculations use
phenomenological matrix elements that should be applicable to a
variety of production processes \cite{ref:leibovich}, including such
kinematically 
different regimes as hadron collisions at the Tevatron collider and
\epem\ collisions at the \FourS.

This paper presents an analysis of \jpsi, \chicone,
\chictwo, and \psitwos\ mesons produced in $B$ decays at the \FourS\
resonance.  \jpsi\ and \psitwos\ mesons are reconstructed in the \epem\ and
\mumu\ decay modes, and the \chicone\ and \chictwo\ in the $\jpsi\gamma$ final
states.  We also reconstruct $\psitwos\to\jpsi\pipi$.

The results include new measurements of previously observed 
decays \cite{ref:cleo, ref:bellechic} as well as 
momentum and helicity
distributions not previously measured.

\section{The BaBar Detector and the PEP-II Collider}
\label{sec:babar}

The \babar\ detector is located at the \pep2\ \epem\ storage rings
operating at the Stanford Linear Accelerator Center.
At PEP-II, 9.0\gev\ electrons collide with 3.1\gev\ positrons to produce
 a center-of-mass energy of 10.58\gev, the mass of the \FourS\ resonance.

The \babar\ detector is described elsewhere~\cite{ref:babar}; here we give 
only a brief overview.
Surrounding the interaction point is a 5-layer double-sided
silicon vertex tracker (SVT), 
which provides precision spatial information for
all charged particles, and also measures their energy loss (\dedx).
The SVT is the primary detection device for low momentum
charged particles. Outside the SVT, a 40-layer drift chamber (DCH) 
provides measurements of the transverse momenta \pt\ of charged particles
with respect to the beam direction. 
The resolution of the \pt\ measurement for tracks with momenta above 1 \gevc 
is parameterized as 
\begin{equation}
\frac{\sigma(\pt)}{ \pt} = 0.13\cdot\pt\% + 0.45\%,
\end{equation}
where \pt\ is measured in \gevc.
The drift chamber also measures
\dedx\ with a resolution of 7.5\%.
Beyond the outer radius of the DCH is a detector of internally reflected 
Cherenkov radiation (DIRC), which is used primarily for charged hadron
identification. The  detector consists of quartz bars in
which Cherenkov light is produced as relativistic charged particles traverse
the material. The light is internally reflected along the length of
the bar into a water-filled  
stand-off box mounted on the rear of the detector.  The Cherenkov
rings expand in the stand-off box and  
are measured with an array of photomultiplier tubes mounted on its
outer surface. A CsI(Tl) crystal  
electromagnetic calorimeter (EMC) is used to 
detect photons and neutral hadrons, 
as well as to identify electrons.  The energy resolution of the
calorimeter is parameterized as:
\begin{equation}
\frac{\sigma(E)}{E} = \frac{2.3\%}{E^{\frac{1}{4}}} \oplus 1.9\%,
\end{equation}
where the energy $E$ is measured in \gev.
The EMC is surrounded by a superconducting 
solenoid that produces a 1.5-T magnetic field. The instrumented flux return
(IFR) consists of multiple layers of resistive plate chambers (RPC)  
interleaved with the flux return iron. 
The IFR is
used in the identification of muons and neutral hadrons.

Data acquisition is triggered with a two-level system.  
The first level uses fast algorithms implemented in hardware that examine
tracks in the DCH and energetic clusters in the EMC.
The second level retains events in which 
the track candidates point back 
to the beam interaction region, 
or in which the EMC cluster candidates are correlated in time
with the rest of the event and exceed the energy of a minimum ionizing
particle. 
Over 99.9\% of \BB\ events pass the second level trigger.
A fraction of all events that pass
the first level are passed through the second
to allow monitoring of its performance.

\section {Coordinate System and Reference Frames}

We use a right-handed coordinate system with the $z$ axis along the electron
beam direction and 
the $y$ axis upwards, with origin at the nominal beam
interaction point.  The polar angle $\theta$ is measured from the $z$
axis and the azimuthal angle $\phi$ from the $x$ axis. 
Unless otherwise stated, kinematic quantities are 
calculated in the rest frame of the detector.  The other reference frame we 
commonly use
is the center of mass of the colliding electrons and positrons, which 
we call the center-of-mass frame.

\section{Event Selection}\label{sub:evtsel}
     
The data used in these analyses 
were collected
between October 1999 and October 2000 and 
correspond to an integrated 
luminosity of 20.3\invfb\ taken 
on the $\FourS$ and 2.6\invfb\ taken off-resonance at an energy 0.04\gev\  
lower than the peak, which is below the threshold for \BB\ production
and therefore includes only continuum processes. 

We use an equivalent luminosity of simulated data \cite{ref:geant}, 
including
both \BB\ and 
continuum, to study the efficiency of the analysis.

We require events to satisfy 
criteria that are intended to have high efficiency for \FourS\
events while rejecting a significant fraction of continuum events and 
strongly suppressing beam gas events. The event must
satisfy either the DCH components of both trigger levels
or the EMC components of both, and have three or more high-quality
tracks in the angular  
region with full tracking acceptance, $0.41 <
\theta < 2.54$. 
Reconstructed charged particles are considered high-quality tracks if
they have at least 12 hits in the DCH, $p_T>
100$\mevc, and a point of closest approach to the beam spot of 
$<3$\cm\ in $z$ and $<1.5$\cm in $xy$.

\begin{figure}
\includegraphics[width=\linewidth]{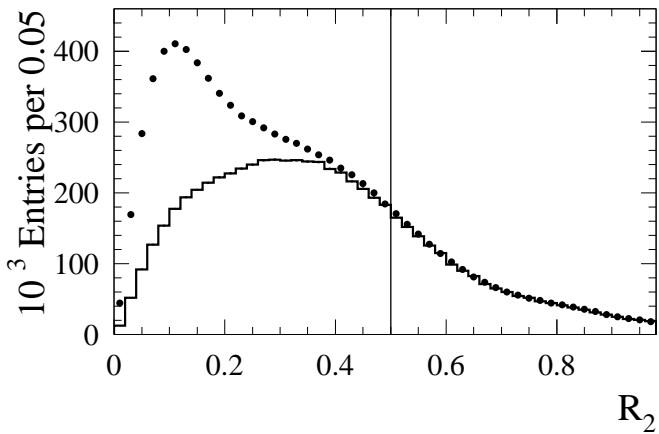}
\caption{ 
$R_2$ distribution after all other selection criteria have been
applied.  Points are on-resonance data, histogram is 
off-resonance data scaled to the same luminosity.  The
vertical line denotes the requirement imposed on $R_2$.
}
\label{fig:r2plot}
\end{figure}

The ratio of the second to the zeroth Fox-Wolfram moment 
$R_2$ \cite{ref:fox} measures
how uniformly the energy in the event is distributed, 
distinguishing the more spherical $\BB$ events with small $R_2$ from
the more jet-like continuum events at large $R_2$
(Fig.~\ref{fig:r2plot}). We require $R_2 < 0.5$. 

The primary event vertex is obtained from all charged particles with  $0.41 <
\theta < 2.54$.  Particles contributing a large \chisq\ are removed
from the vertex and the process is iterated until stable. 
To reject events due to a beam particle
striking the beam pipe or a residual gas molecule, 
we require the primary event vertex to be within 0.5\cm\ of the beam
spot in 
$xy$ and 6\cm in $z$.

Finally, the event must include total visible energy (charged particles plus
unassociated EMC 
clusters above 30\mev) greater than 4.5\gev.

The efficiency for simulated $\FourS\to\BB$ events is 95.4\%.  More
relevant is the ratio $C_E$ of
this efficiency to that for events containing the charmonium decay of
interest.  $C_E$ is calculated for each of the final states using
simulated data.  Inaccuracies in the simulation 
of the tracking systems produce
an uncertainty of 1.1\%, common to all modes.

\subsection{Determination of Number of \FourS\ Mesons}

The number of \FourS\ events satisfying the 
above selection criteria
($N_\Upsilon$) is obtained from the total number of events satisfying
the criteria by subtracting the component due to the continuum.
\begin{equation}
N_\Upsilon = N_{on} - \kappa \cdot R_{off} \cdot M_{on},
\label{eq:bcount}
\end{equation}
where
\begin{description}
\item[$N_{on}$] is the number of events satisfying the criteria in the
on-resonance data set;
\item[$M_{on}$] is the number of muon pairs in the on-resonance data
set;
\item[$R_{off} = N_{off}/M_{off}$] is the ratio of the 
number of events satisfying the
hadronic selection criteria to the number of muon 
pairs in the off resonance data; and
\item[$\kappa$] $= 1.0000\pm 0.0025$ allows for differences in the
ratios of continuum and muon-pair cross sections and efficiencies
between
on-resonance and off-resonance data.
\end{description}

The two
highest-momentum tracks in muon pair events must both deposit less
than 1\gev\ in the EMC 
and satisfy $|\cos\theta|<0.7485$ in the 
center-of-mass frame (to be within the region of full tracking
efficiency).  The tracks must be within $10^\circ$ of being
back-to-back, and must have a combined mass greater than
7.5\gevcc. 
Since the number of muon pairs in the on and off resonance data sets
appears only as a ratio in Equation~\ref{eq:bcount}, $N_\Upsilon$
depends only weakly on the details of the selection criteria. 
Varying these over reasonable ranges changes
$N_\Upsilon$ by 0.5\%, which we take as a systematic error. 
The on-resonance data contains 7.8 times as many muon
pairs as the off-resonance data. 

Changes to the trigger configuration caused 
$R_{off}$ to vary from 4.89 to 4.94 during the period in which the
data was collected.
For the purposes of this calculation, the data is grouped into periods
of compatible $R_{off}$.  Combining the data in a single group changes
$N_\Upsilon$ by 0.3\%, which is taken as a systematic error.
In principle, $R_{off}$ could also vary due to beam gas backgrounds.
However, 
the $z$ distribution of the primary event vertex
indicates that less than 0.1\% of events are due to beam gas.  

We find $N_\Upsilon = (21.26\pm0.17)\times 10^6$.  The 0.8\% 
uncertainty is systematic and includes the 0.5\%, 0.3\% and 0.1\%
contributions from muon pairs, 
$R_{off}$, and beam gas described above.
The largest component
is from the 0.25\% uncertainty on $\kappa$, which corresponds to a
0.6\% uncertainty on $N_\Upsilon$. 

\begin{figure}
\includegraphics[width=\linewidth]{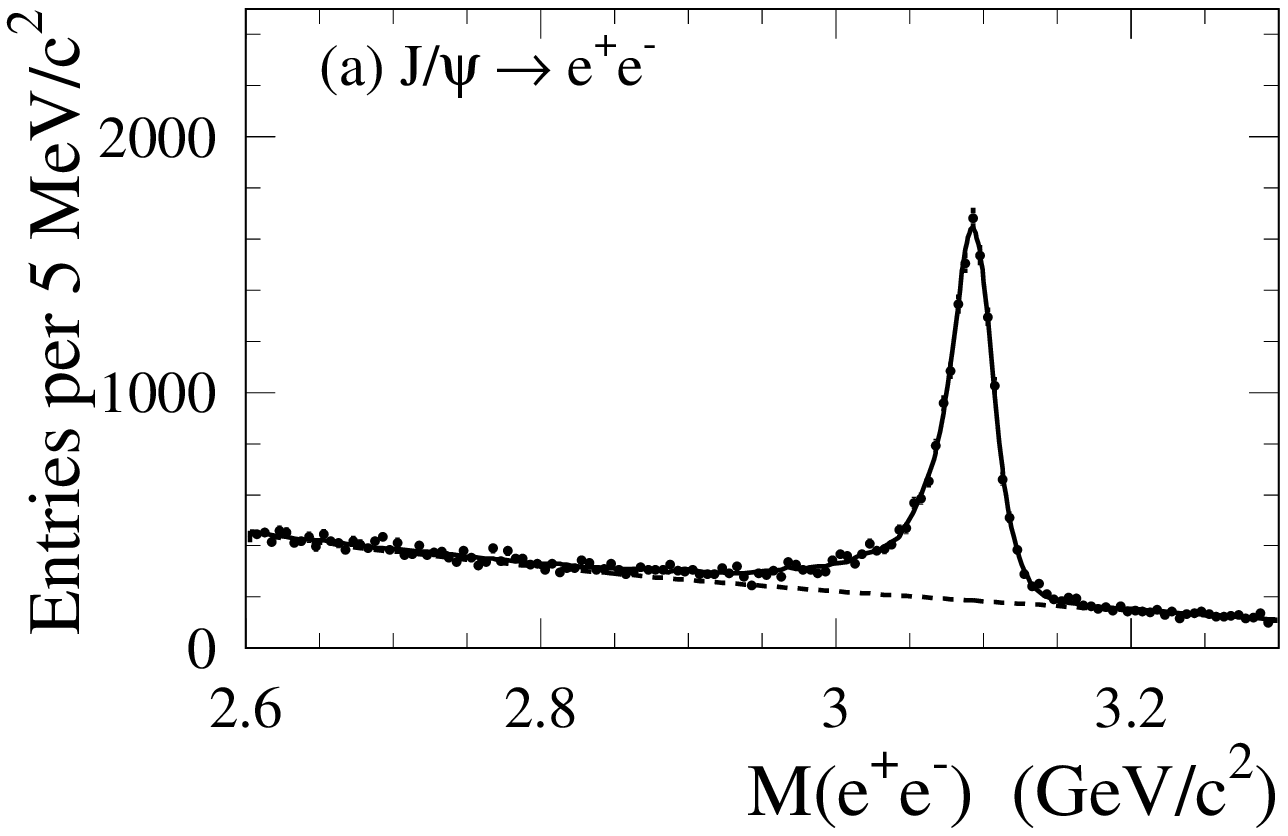}
\includegraphics[width=\linewidth]{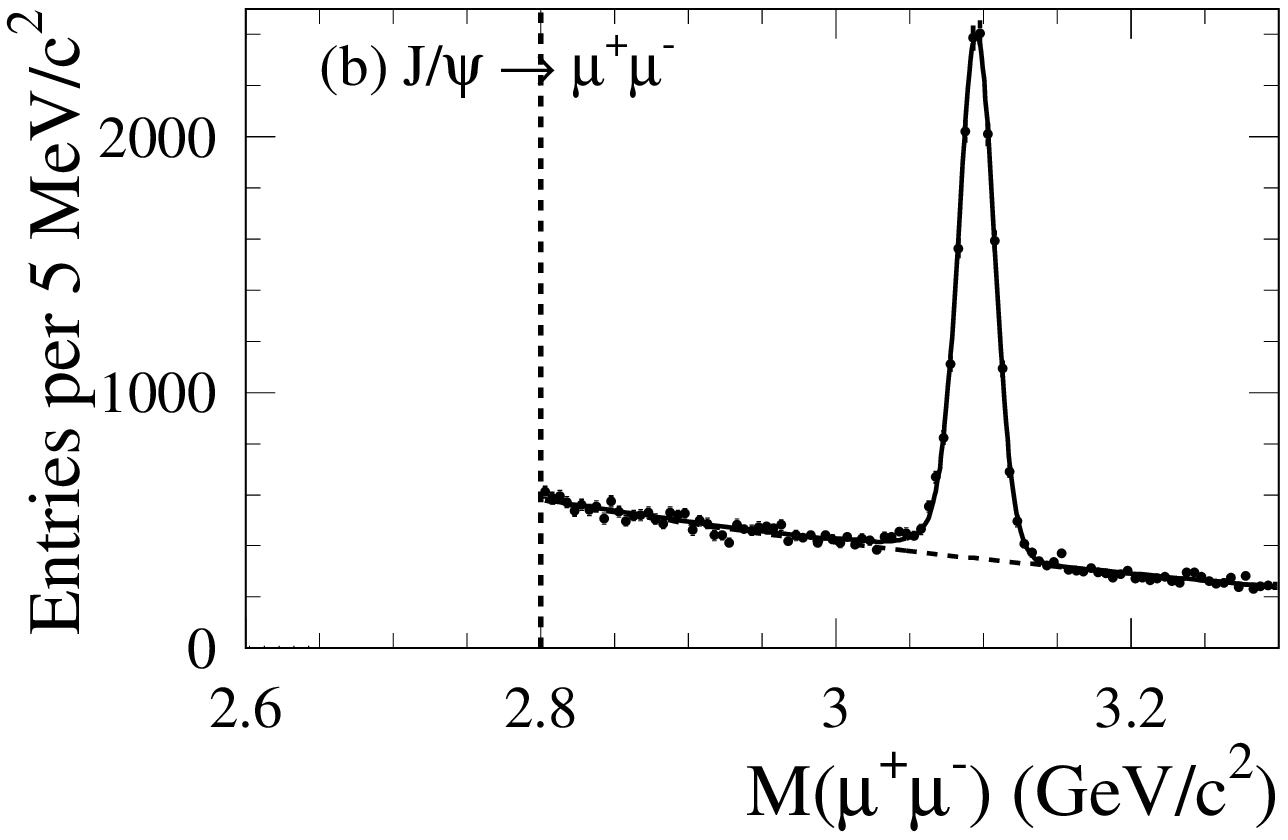}
\caption{Mass distribution of 
\jpsi\ candidates reconstructed in the (a) \epem\ and (b)
\mumu\ final states. The vertical dashed line in (b) marks the lower
edge of the mass range used in the \mumu\ final state.
}
\label{fig:jpsimass}
\end{figure}

\section{\jpsi\ Production}

\subsection{\jpsi\ Reconstruction}\label{sub:jpsi}

We reconstruct \jpsi\ candidates in selected events using the \epem\
and \mumu\ final states.  The leptons are required to
satisfy the
track quality criteria listed earlier.  
Electron candidates are further required to have a distance
of closest
approach to the beam line of less than 0.25\cm\ to reject electrons
produced by photon conversions.  
 
Both leptons are required to fall in the
angular range $0.410 <\theta<2.409$\rad\  
(the overlap of the SVT and EMC
coverage). 
Simulation indicates that $(75.3\pm0.9)$\% of \jpsi\ decays give both
leptons in this region.
As described in Sec.~\ref{sub:pstar}, the \jpsi\ momentum distribution
in the simulation \cite{ref:jetset} is slightly different from the observed
distribution. This difference, and a small variation of efficiency
with momentum, produce the uncertainty in the efficiency.

We obtain the efficiency for the leptons to satisfy the quality criteria 
by comparing the performance of the 
independent SVT and DCH
tracking systems in hadronic events.  
The corresponding uncertainty in the efficiency is 2.4\% per \jpsi.

One particle in a \jpsiee\ candidate must satisfy the ``very tight''
electron identification criteria described
below.  The other must satisfy the
``tight'' criteria displayed in square brackets.
\begin{itemize}
\item Difference between measured and expected energy loss in the DCH
between 
$-2\sigma$ and $+4\sigma$, where $\sigma$ is the measurement error 
[$-3\sigma$ and $+7\sigma$];
\item Ratio of energy measured in EMC to measured momentum $E/p$ in
range 0.89 to 1.2  [0.75 to 1.3];
\item Associated EMC cluster must include at least four crystals [same];
\item Lateral energy distribution LAT \cite{ref:lat} of EMC cluster in range
0.1--0.6 [0.0--0.6]; 
\item The $A_{42}$ Zernike moment \cite{ref:zern} of the EMC cluster
$< 0.42$ [no requirement]; and 
\item DIRC Cherenkov angle within $3\sigma$ of expected
value [no requirement].
\end{itemize}

LAT is a measure of the radial energy profile of the cluster, and is
used to suppress clusters from electronic noise (very low LAT) or hadronic
interactions (high LAT). 
$A_{42}$ measures the azimuthal 
asymmetry of the cluster about its peak, distinguishing
electromagnetic from hadronic showers. 

We reduce the impact of bremsstrahlung by combining photons radiated
by electron candidates with the track measured in the tracking 
system (``bremsstrahlung-recovery'').  Such photons must have
EMC energy greater 
than 30\mev\ and a polar angle within 35\mrad\ of the electron
direction. The azimuthal angle of the photon must be 
within 50\mrad\ of the electron
direction at the beamspot
or be between this direction and 
azimuthal location of the electron shower in the EMC.  

For $\jpsi\to\mumu$ candidates, 
one muon candidate must satisfy ``tight'' criteria, while
the other satisfies ``loose'' criteria (shown in square brackets):
\begin{itemize}
\item Energy in calorimeter between 0.05 and 0.40\gev [$<0.50$\gev];
\item Number of IFR layers $N_{IFR}\ge 2$ [same];
\item Particle penetrates at least 2.2 interaction lengths 
$\lambda$ of detector material [$2\lambda$]; 
\item Measured penetration within $\pm 0.8 \lambda$ of the value
expected for that momentum [$\pm 1.0 \lambda$];
\item An average of less than 8 hit strips per IFR layer [same];
\item The RMS of the hits per IFR layer less than 4 [same];
\item For candidates in the forward endcap, the number of hit 
IFR layers
divided by the total number of layers between the first and last hit
layers must be $>0.34$ to reject beam background in the outermost
layer [$>0.30$];
\item $\chi^2$ of the track fit in the IFR $<3\times N_{IFR}$ [same]; and
\item $\chi^2$ of match between track from SVT and DCH and that found
in the IFR $<5\times N_{IFR}$ [same].
\end{itemize}

The particle identification efficiencies are obtained by comparing the
yield of \jpsi\ mesons applying various criteria to one or both
tracks. 
The efficiency for \jpsiee\ satisfying the angular acceptance and
track quality criteria 
is 90.5\% with a systematic error
of 1.8\%.  For \jpsimm, it is 71.7\% with a systematic error of 1.4\%. 
The systematic errors are somewhat conservative, in that they include
a component due to the \jpsi\ statistics. 
Misidentification of hadrons as muons 
is higher than misidentification as electrons, producing \jpsi\
background levels that are approximately a factor of two higher.

Finally, the \jpsi\ candidate must have momentum in the center of mass
$\pstar<2.0$\gevc.  This requirement is fully efficient for a \jpsi\
meson from $B$ decay but rejects approximately 74\% of those
produced in the continuum \cite{ref:cont}.

More than one \jpsi\ candidate may be found in an event. A second
candidate is observed in 0.8\% of events that include at least one.

\subsection{Extraction of Number of \jpsi\ Mesons}

The mass of the \jpsi\ candidate
is obtained after constraining the two leptons
to a common vertex.  In less than 1\% of events, the vertex fit
does not converge, and we instead use four-vector addition to obtain
the candidate mass.  The mass resolution is poorer by approximately
1\% for these events. Figure~\ref{fig:jpsimass} shows the mass
distribution of the selected candidates in the two lepton modes. 

The number of \jpsi\ mesons in the mass window used in the fit
(2.6--3.3\gevcc\ for the \epem\ mode, 2.8--3.3\gevcc\ for \mumu)
is
determined by a binned likelihood
fit to the distribution.  The background is
represented by a third-order Chebychev polynomial.  
The probability
distribution function (pdf) for the \mumu\ signal 
is the distribution from simulation convolved with 
a Gaussian distribution, with
mean (allowing for a systematic shift between simulation and
data) and width (allowing for poorer resolution)
free to float in the fit.  The additional smearing is required because
the simulation underestimates the 
actual amount of material in the detector.
The fit returns an offset of 3\mevcc\ and an
additional resolution of 7.8\mevcc.  The
total mass resolution in data is 
approximately 12\mevcc. The simulation includes final state
radiation \cite{ref:photos}
and predicts that $(2.7\pm0.1)$\% of reconstructed \jpsimm\ candidates
fall outside the mass range used in the fit.

To include the impact of bremsstrahlung, the \jpsiee\ signal pdf
includes four components, corresponding to mesons 
where neither electron has undergone bremsstrahlung or at least one,
and mesons for which the bremsstrahlung-recovery 
process has located a photon or not.  The shapes are derived from simulated
data, but 
the relative weights of three of the
four components
are allowed to float in the fit.  The fraction of events that did
not undergo bremsstrahlung but had a photon assigned by the
bremsstrahlung-recovery process is nominally fixed to the value
predicted by 
simulation, although it is varied to obtain a systematic error.

An estimated
$(7.3\pm0.8)$\% of reconstructed \jpsiee\ candidates
fall outside the mass
range, which can be compared to the value of 6.1\% 
if the relative pdf weights are fixed to the values predicted 
by simulation.   

We perform similar fits to the off-resonance data with all signal fit
parameters fixed except for the number of mesons.  
The result is
scaled by the ratio of on- to off-peak luminosity 
and subtracted to obtain the number of
mesons attributable to $B$ decay, which appears in
Table~\ref{tab:yield} as the net meson yield.

The fitting procedure is validated and systematic errors on its
results are obtained by comparing the generated number of events with
the fit number of events for many simulated mass distributions
convolved with a Gaussian distribution.  For the purposes of this
test, we increase the statistics of
the simulated data by relaxing  the particle identification and
track-quality requirements. 
We also test second and fourth-order Chebychev
polynomials for the background pdf.  We perform these tests for the \chicone,
\chictwo, and \psitwos\ mass distributions as well.

We vary fit parameters that are  fixed during the fit
to obtain an additional systematic contribution.  In the
case of the \jpsi, we 
vary the bremsstrahlung-recovery error rate from one-half to twice its
nominal 
value. The systematic errors on the fit yields are 0.7\% for \jpsimm\
and 4.1\% for \jpsiee.
  
\subsection{Determination of the $B\to\jpsi X$ Branching Fraction}

We calculate values for the  $B\to\jpsi X$ branching fraction using
the \epem\ and \mumu\ final states separately, then combine the two.
The equation for the branching fraction is the same for both cases (and
for the other mesons studied):
\begin{equation}
\BR = \frac{ N_\psi}{2\cdot N_\Upsilon \cdot \epsilon_C 
   \cdot {\cal B}_c} \cdot
  C_E,
\label{eq:bfform}
\end{equation}
where 
\begin{description}
\item[$N_\psi$] is the net number of mesons in the mass fit range 
after continuum subtraction;
\item[$\epsilon_C$] is the efficiency for a meson
to satisfy the selection criteria, including the requirement that the
mass fall in the mass fit range;
\item[${\cal B}_c$] is world average \cite{ref:pdg2000} for the
relevant secondary charmonium 
branching fraction. For the \jpsi, this is for \jpsiee\ or
\jpsimm;
\item[$C_E$] corrects for the difference in event selection efficiency
(Sec.~\ref{sub:evtsel}) 
between generic \BB\ events and charmonium events. It 
is equal to the efficiency for generic \BB\ events divided by the
efficiency for the relevant charmonium final state.
\end{description}

Table~\ref{tab:yield} summarizes the meson yields and efficiencies.
It also presents the branching fraction product $\BR\cdot {\cal B}_c$, an
experimental quantity that does not depend on the secondary charmonium
branching fractions.
There is a 3.1\% systematic error common to both 
modes---and, in fact, to all final states we study---due to acceptance
(1.2\%), track 
quality selection (2.4\%), uncertainty on $C_E$ (1.1\%), and number of
\FourS\ (0.8\%).  

The separate \epem\ and \mumu\ branching fraction
measurements are averaged to obtain the
final result (Table~\ref{tab:sum}).  Each measurement is weighted in
the average by the 
inverse of the square of the statistical error plus the square of the
systematic errors unique to that mode. 
The common systematic error is the largest component of the 
$\BR(B\to\jpsi X)$ uncertainty.

\begin{table*}
\caption{Meson yield in on-resonance (20.3\invfb) and off-resonance
(2.6\invfb) data, and net yield after 
continuum subtraction. $\epsilon_C$ and $C_E$ are the meson
reconstruction efficiency and the relative event selection efficiency; 
$S_N$ and $S_\epsilon$ are the systematic errors on the meson yield
and reconstruction efficiency; {\em Stat} is the statistical
error. The systematic errors on the branching fraction products
include components unique to that final state.  The total uncertainty
values 
include those listed in the ``Common'' rows. ${\cal B}_c$ are the secondary 
branching fractions and $S_B$ the uncertainty.  {\em 
Tot} is the total systematic error (percentage) on the $B$
branching fraction to that final state.}
\begin{tabular}{|l|ccc|cc|ccc|ccc|cc|c|}
\hline \hline
 & \multicolumn{3}{c|}{ Meson Yield $N_\psi$} 
  & \multicolumn{2}{c|}{Efficiencies} & \multicolumn{3}{c|}{Uncertainties (\%)}
 & \multicolumn{3}{c|}{\BR\ Product $\times 10^6$} & & & \\
\em Mode & \em On & \em Off & \em Net & $\epsilon_C$ & $C_E$ &  $S_N$ & $S_\epsilon$ &  
     \em Stat  & \em Value &\em Stat &\em Sys & ${\cal B}_c$ & $S_B$
      \% & \em Tot \% \\
\hline
$\jpsi\to\ellell$ & & & & & & & & & & &   & & &  \\
  \hspace{0.1in} \epem  & $16095 \pm 242$ & $23 \pm 15$ & $15914 \pm 268$
   & 0.589 & 1.02 & 4.1 & 2.0 & 1.7 
   & 650 & $\pm 11$ & $\pm 30$ & 0.0593 & 1.7 & 4.9 \\
  \hspace{0.1in} \mumu & $13683\pm 154$ & $67 \pm 18$ & $13159 \pm 210$ 
   & 0.500 & 0.99 & 0.7 & 1.4 & 1.6 
   & 615 & $\pm 10$ & $\pm 10$ & 0.0588 & 1.7 & 2.3 \\  
 \hspace{0.1in} Common  & --  & -- & -- 
   & --&    --     &  --  &  3.1  &  --
   & -- & -- & 3.1\%  &  -- & 0. & 3.1\\
\hline 
$\chicone\to\jpsi\gamma $  & & & & & & & & & & &   & & &  \\
  \hspace{0.1in} \jpsiee  &  $512 \pm 62$  & $-2 \pm 3$     & $528 \pm 67$  
  & 0.191 & 1.04  & 6.4 & 3.1 &12.7
  & 68 & $\pm  9$ & $\pm 5$ & 0.0593 &1.7 &7.3\\
  \hspace{0.1in} \jpsimm  &  $614 \pm 72$  &   $3 \pm  6$   & $592 \pm 86$  
  & 0.201 & 1.00  & 5.4 & 2.0  &14.5 
  & 69 & $\pm 10$ & $\pm 4$ & 0.0588 & 1.7  & 6.0\\
 \hspace{0.1in} Common  & --  & -- & -- 
   &    -- &   --   &  -- &  4.5 &  --
  & -- & -- & 4.5\%  &  0.316  & 10.1  & 11.1\\
\hline 
$\chictwo\to\jpsi\gamma $ & & & & & & & & & & &   & & &  \\
  \hspace{0.1in} \jpsiee  &  $168 \pm 48$ &   $-5 \pm  3$      &  $210 \pm 54$  
  & 0.197   & 1.04    & 3.9     & 9.6       & 25.7 
  & 26 & $\pm 7$ & $\pm 3$ & 0.0593 & 1.7     & 10.3\\
  \hspace{0.1in} \jpsimm  & $208 \pm 54$  & $4 \pm 5$     & $174 \pm 66$ 
   & 0.207 & 1.00  & 10.4    & 12.1        & 38.0 
  & 20 & $\pm 8$ & $\pm 3$  & 0.0588  & 1.7    & 16.0\\
 \hspace{0.1in} Common  & --  & -- & -- 
   & --   &--      &--   & 5.3        &-- 
  & -- & -- & 5.3\% & 0.187 & 10.7 & 11.9 \\  
\hline 
$\psitwos\to\ellell$  & & & & & & & & & & &   & & &  \\
  \hspace{0.1in} \epem  &  $573 \pm 52$   & $-6 \pm 8$    & $623 \pm 81$  
  & 0.594 & 1.05  &  4.3  & 1.9   &  12.9
  &  25.8 & $\pm 3.3$ & $\pm 1.2$ & 0.0078 &  --  &  -- \\
  \hspace{0.1in} \mumu  &  $437 \pm 44$  &   $5 \pm 10$   & $400 \pm 92$  
   & 0.535 & 1.01  & 3.1   &  1.6   &  23.0 
   & 17.8 & $\pm 4.1$ & $\pm 0.5$ & 0.0067 &  --  & --  \\
   \hspace{0.1in} Common  & --  & -- & -- 
   &   --   & --       &   -- & 3.1   & -- 
  & -- & -- & 3.1\% &  --   & --   &   3.1\\ 
\hline 
\multicolumn{2}{|l}{$\psitwos\to\jpsi\pipi$}  & & & & & & & & & & &   &  &  \\
  \hspace{0.1in} \jpsiee  &   $474 \pm 43$   &   $0 \pm  2$    & $476 \pm 45$  
   & 0.205   & 0.99   & 2.7    & 2.1       & 9.5 
  & 53.9 & $\pm 5.1$ & $\pm 1.8$  & 0.0593 & 1.7    & 3.8\\
  \hspace{0.1in} \jpsimm  &  $493 \pm 42$   &   $0 \pm  3$   & $496 \pm 47$   
  & 0.215  & 0.98  & 2.3    & 1.4       &  9.5
  & 53.3 & $\pm 5.1$ & $\pm 1.4$  &  0.0588 & 1.7    & 3.2\\
  \hspace{0.1in} Common  & --  & -- & -- 
  & --   & --    & --  &  3.5    & --
  & -- & -- & 3.5\%  &  0.305 & 5.2    &   6.3\\
\hline 
\hline
\end{tabular} 
\label{tab:yield}
\end{table*}

\section{\chic\ Production} \label{sec:chic}

We reconstruct \chicone\ and \chictwo\ mesons in the $\jpsi\gamma$
decay mode.  The \jpsi\ mesons must satisfy the criteria listed in
Sec.~\ref{sub:jpsi}, with the additional requirement that the
\jpsi\ candidate mass $m$ satisfy $3.05<m<3.12$\gevcc\ for \epem\
decays and 
$3.07<m<3.12$\gevcc\ for \mumu.  
An estimated $(74.0\pm0.4)$\% of \jpsiee\ and
$(91.4\pm0.3)$\%  of \jpsimm\ mesons that satisfy all other criteria  
satisfy this additional mass selection. 

Photon candidates are EMC clusters in the angular range
$0.41<\theta<2.409$\rad\  
with energy between 0.12\gev\ and 1.0\gev.  Hadronic
showers are suppressed by requiring
LAT less than 0.8 and $A_{42}<0.15$, while clusters from nearby
hadronic showers are suppressed by requiring that candidates
be at least 9\degrees\ from all charged tracks.

Most photons satisfying these requirements are produced in \piz\
decay.  We reject a candidate that, when combined with any other photon,
produces a mass between 0.117\gevcc\ and
0.147\gevcc. The second photon must have energy greater than 30\mev
and LAT$<0.8$, with no requirement on $A_{42}$ or distance from charged
tracks. 

\begin{figure}
\includegraphics[width=\linewidth]{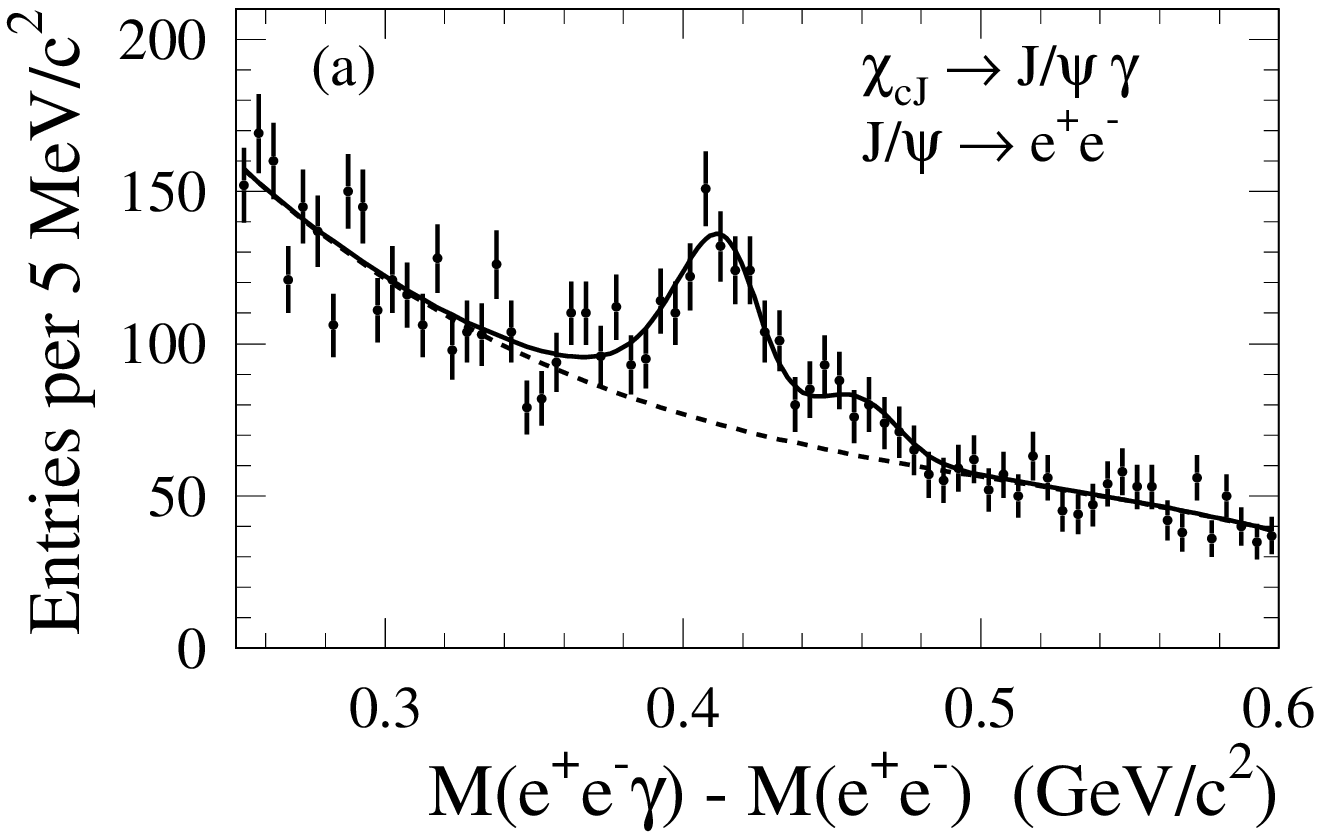}
\includegraphics[width=\linewidth]{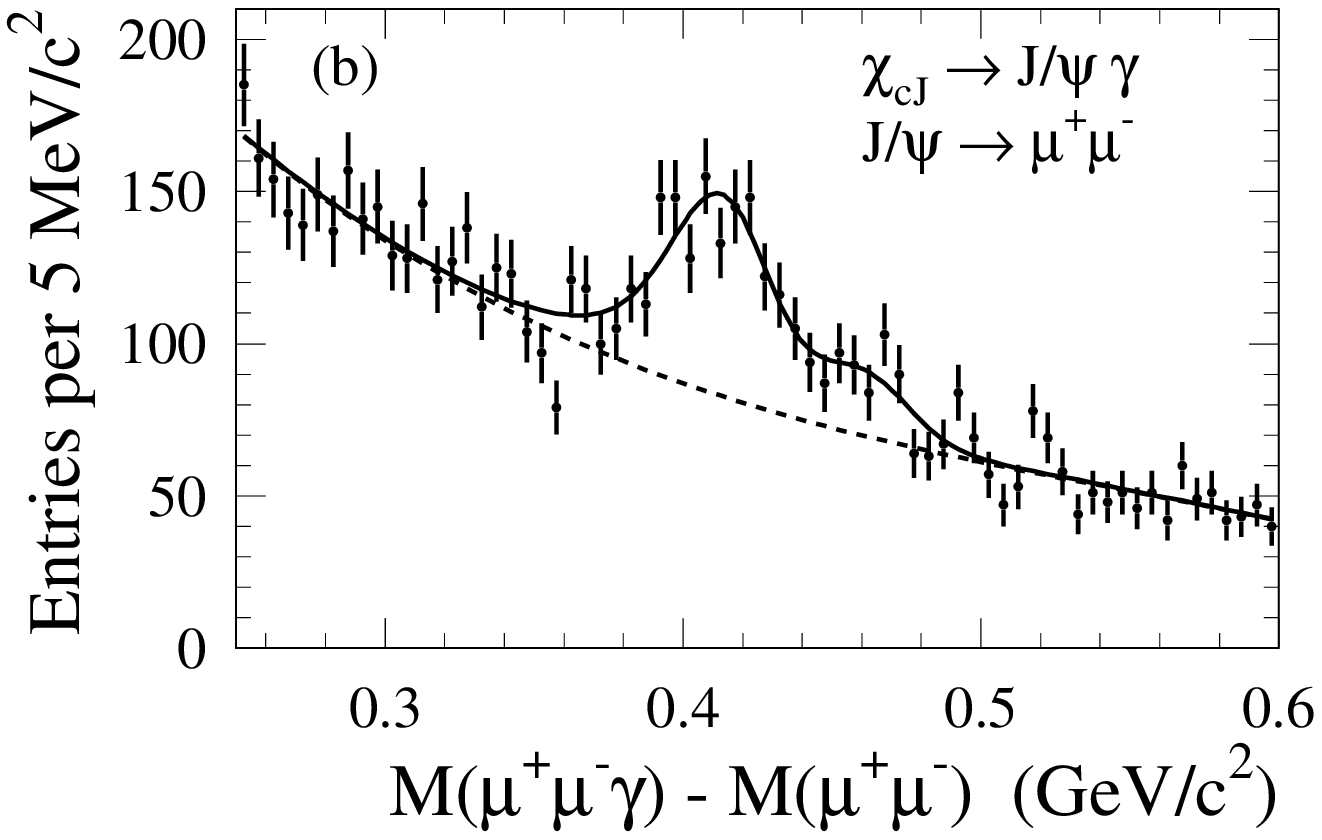}
\caption{\chicone\ and \chictwo\ candidates reconstructed in the
$\jpsi\gamma$ final state.  Mass difference between the $\jpsi\gamma$
and \jpsi\ candidates when the \jpsi\ is reconstructed in 
the (a) \epem\ and (b) \mumu\ final states.}
\label{fig:chicdmass}
\end{figure}

A systematic error due to the photon selection criteria is
obtained by comparing the branching ratio
$\tau^+ \to h^+ \piz \piz / \tau^+ \to h^+ \piz$
in data to that in simulation, where $h^+$ is any charged track.
We also
vary the minimum energy requirement from 0.10 to 0.14\gev\ and 
test alternative \piz\ veto regions. We obtain a systematic error
of 3.1\% for the \chicone\ and 4.4\% for the \chictwo\ common to both
the \epem\ and \mumu\ final states.  
An additional component to the systematic error arises from changes in
the shape of the background, shown in Fig.~\ref{fig:chicdmass},
affecting the 
fit results.  It
is
specific to each final state and mode, and amounts to 
2.2\% (\epem) and 1.3\% (\mumu) for
the \chicone, and 9.2\% (\epem) and 11.9\% (\mumu) for the \chictwo. 

The photon is constrained to originate at the \jpsi\ vertex in the
calculation of the \chic\ four-momentum. We require $\pstar<1.7$\gevc,
a requirement that is satisfied by \chicone\ or \chictwo\ mesons from
$B$ decays. 

We determine the number of mesons from a fit to the plot of the mass
difference 
between the candidate and the daughter \jpsi\ masses
(Fig.~\ref{fig:chicdmass}). We use different signal
pdfs for the \chicone\ and \chictwo.  These are formed by convolving
the pdf calculated by 
simulation with a Gaussian distribution, where the offset and sigma
are constrained to be the same for the \chicone\ and \chictwo.  The
background is described by a third-order Chebychev
polynomial. Systematic errors on the fit are obtained as for the \jpsi.
The correlation coefficient between the number of \chicone\ and
\chictwo\ mesons obtained from the fit is 0.19.

Equation~\ref{eq:bfform} is used to determine inclusive $B\to\chicone X$
and $B\to\chictwo X$ branching fractions separately for \jpsiee\ and
\jpsimm.  In this case, ${\cal B}_c = \BR(\chicj\to\jpsi\gamma)\cdot
\BR(\jpsi\to\ellell)$, where  $\chicj$ is \chicone\ or
\chictwo\ and $\ell$ is $e$ or $\mu$.  
Table~\ref{tab:yield} summarizes
the yields, efficiencies, uncertainties and branching fraction
products 
$\BR(B\to\chicj X)\cdot\BR(\chicj\to\jpsi\gamma)
\cdot\BR(\jpsi\to\ellell)$.
The 4.5\% (\chicone) and 5.3\% (\chictwo)
systematic errors common to both \epem\ and \mumu\ include photon
reconstruction in addition to the \jpsi\ reconstruction items.

As with the \jpsi, the inclusive $B$ branching fractions to the
\chicone\ and \chictwo\ are calculated separately using the \jpsiee\
and \jpsimm\ decays.  The two values are then combined, distinguishing
uncertainties common to both from those unique to a single final
state. 

The branching fraction obtained for the \chictwo\ is comparable to
that for the \chicone\ and is summarized in Table~\ref{tab:sum}.
This result is consistent with a prediction
from a color octet calculation \cite{ref:bodwinc2}, and is in contrast to
the expectation of a null result in a factorization 
calculation \cite{ref:kuhn}.

\section{\psitwos\ Production}

\begin{figure}
\includegraphics[width=\linewidth]{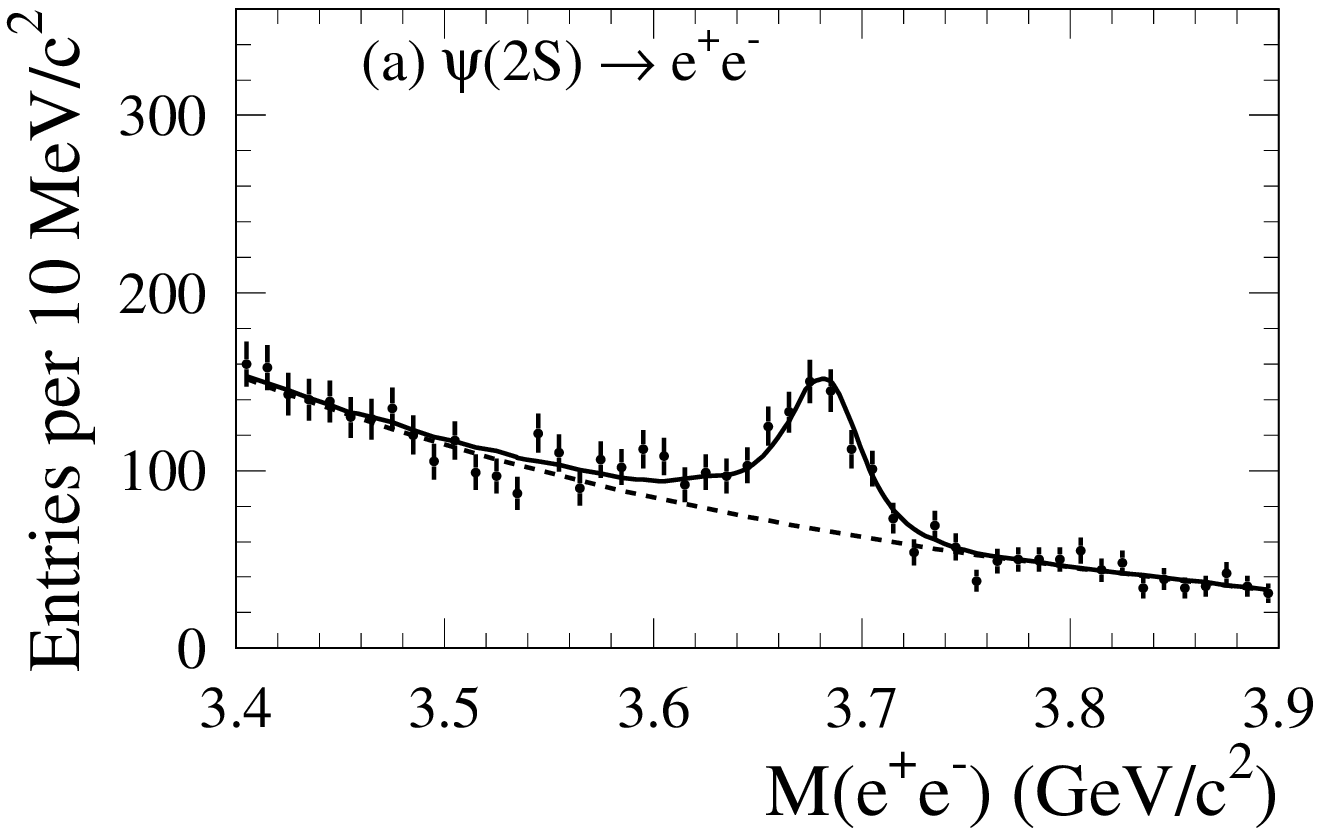}
\includegraphics[width=\linewidth]{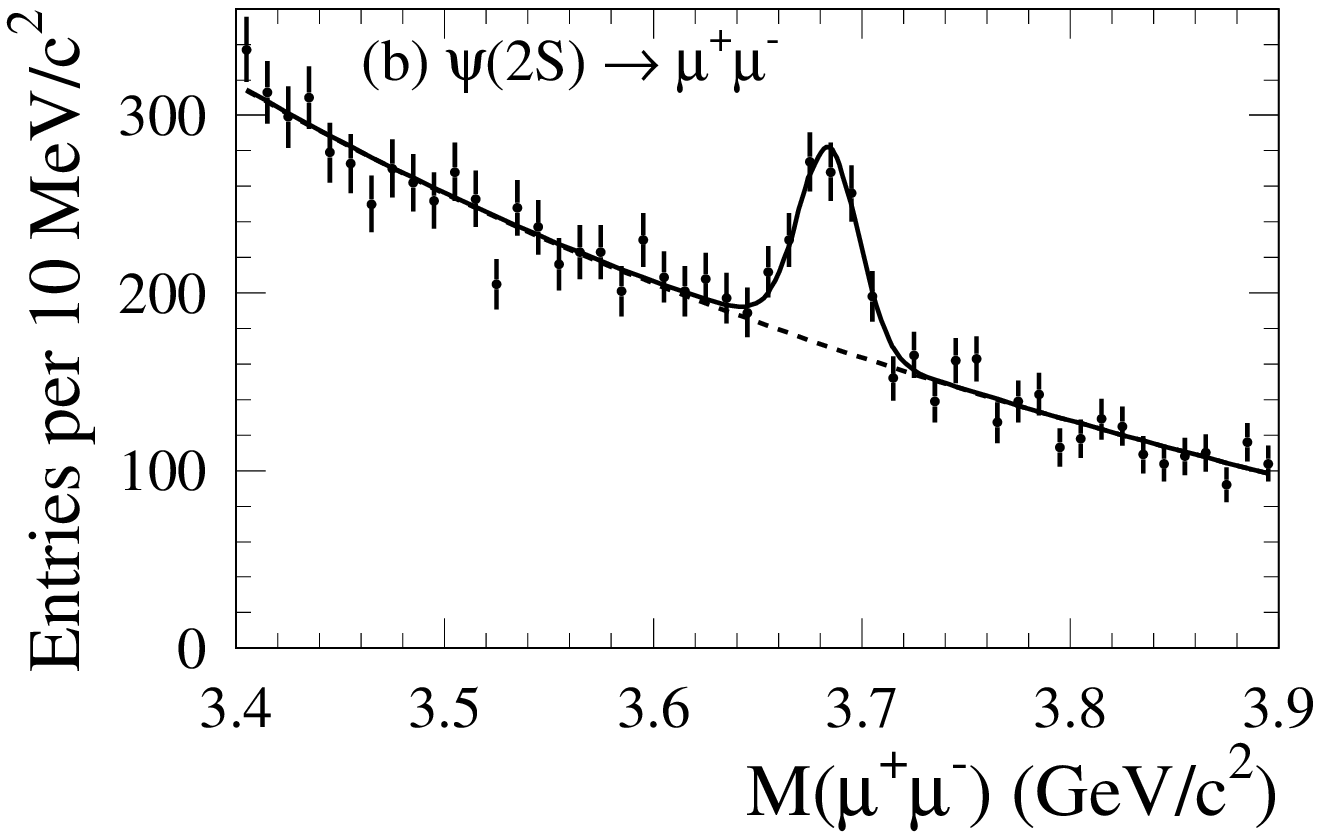}
\caption{Mass distribution of 
\psitwos\ candidates reconstructed in the (a) \epem\ and (b)
\mumu\ final states.}
\label{fig:psi2sllmass}
\end{figure}

The reconstruction of the \psitwos\ in the \ellell\ final state is
very similar to the \jpsi\ reconstruction outlined in
Sec.~\ref{sub:jpsi}, with the \pstar\ requirement tightened to
$\pstar<1.6$\gevc. Figure~\ref{fig:psi2sllmass} shows the resulting
candidate mass distribution. A fit to extract the 
number of mesons in each plot is performed as for the \jpsi, but with the
resolution and bremsstrahlung parameters fixed to the values found in
the higher-statistics \jpsi\ channels.  These parameters are varied
according to their uncertainties as one contribution to the
systematic error on the fit; the remaining contributions are 
determined as for the \jpsi.

These data are used to calculate the branching fraction product 
$\BR(B\to\psitwos X)\cdot \BR(\psitwos\to\ellell)$, 
and are
later used in the determination of the \pstar\ distribution of
\psitwos\ mesons produced in $B$ decay.  However, the extraction of
the $B\to\psitwos X$ branching fraction requires the use of
$\psitwos\to\epem$ and $\psitwos\to\mumu$ branching fractions. Since
this same data set has previously been used to measure these branching
fractions \cite{ref:psi2s}, we do not use $\psitwos\to\ellell$
events to find $\BR(B\to\psitwos X)$.

Instead, we use $\psitwos\to\jpsi\pipi$ for this purpose. 
The reconstruction of a \psitwos\ candidate in this final
state starts with a \jpsi\ candidate satisfying the tighter mass
constraints used in $\chi_{cJ}$ reconstruction.  All charged particles,
including those failing the ``good-track'' criteria, 
are assumed to be pion candidates.  The pion pair is required to be
oppositely charged and to have a mass, calculated by four-vector
addition, in the range 0.45 to 0.60\gevcc.  The mass distribution 
from simulation is compared to 
the measured  \cite{ref:bes} distribution to
obtain a systematic error of 0.5\% on reconstruction efficiency.  
Finally, the \pstar\ of the
\psitwos\ candidate is required to be less than 1.6\gevc.

\begin{figure}
\includegraphics[width=\linewidth]{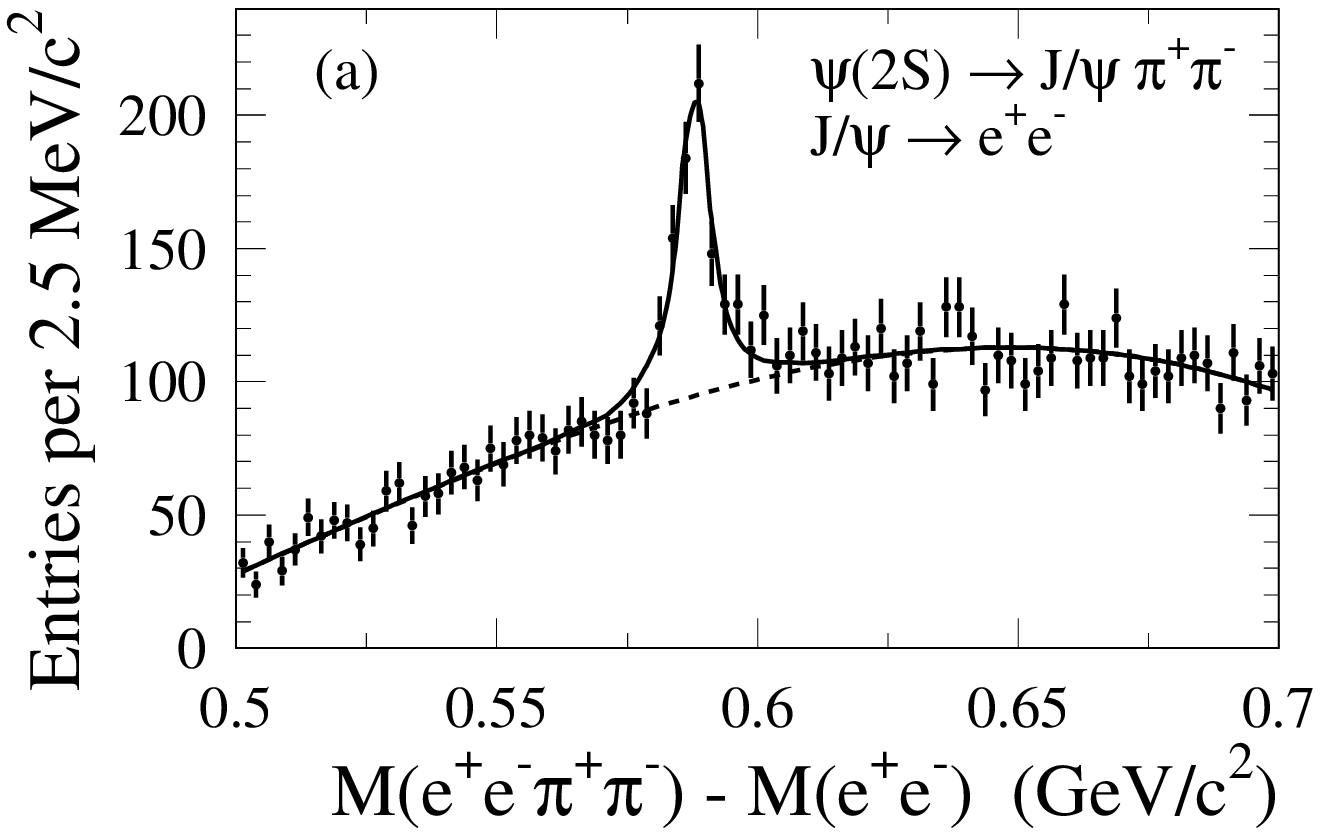}
\includegraphics[width=\linewidth]{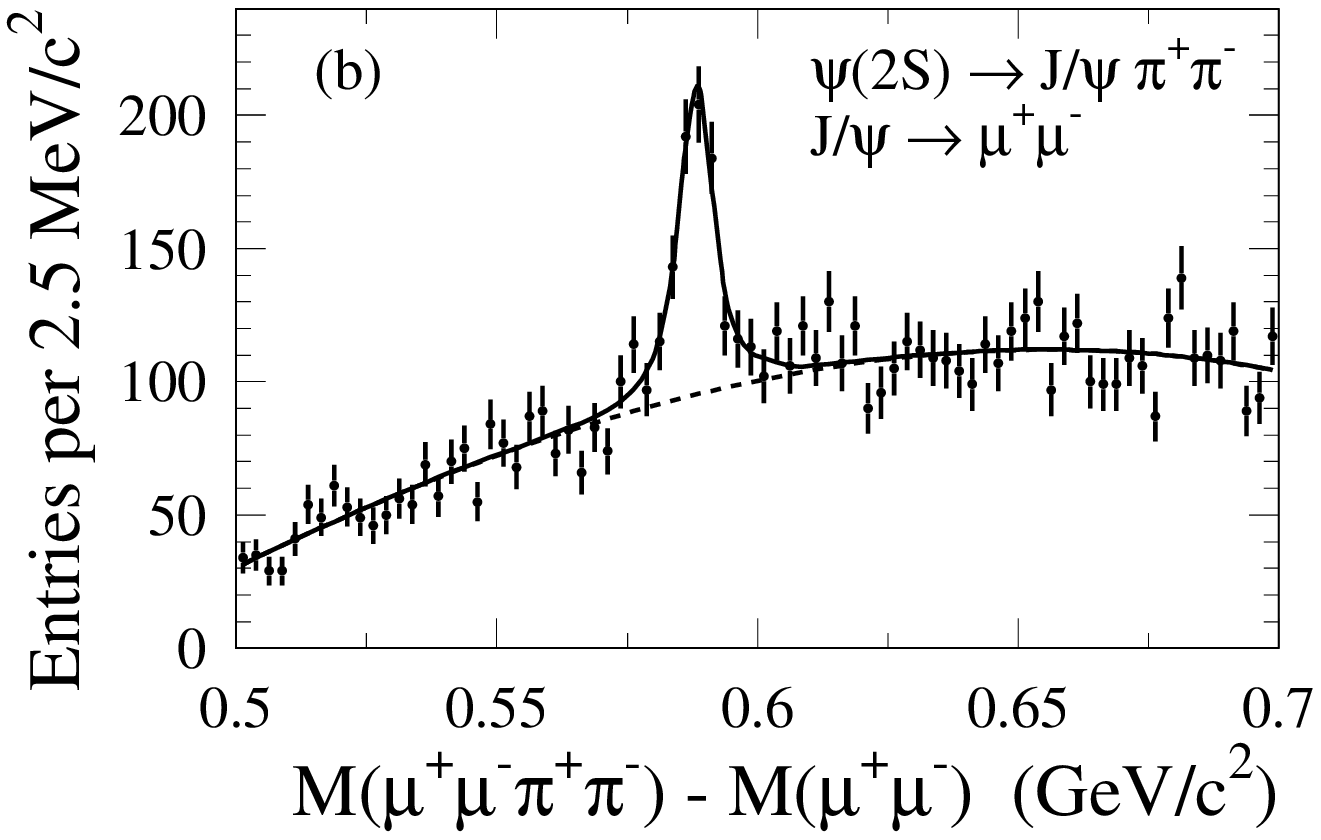}
\caption{\psitwos\ candidates reconstructed in the
$\jpsi\pipi$ final state.  Mass difference between the $\jpsi\pipi$
and \jpsi\ candidates when the \jpsi\ is reconstructed in 
the (a) \epem\ and (b) \mumu\ final states.}
\label{fig:psi2sdmass}
\end{figure}

Figure~\ref{fig:psi2sdmass} displays the mass difference between the
\psitwos\ and the \jpsi\ candidates separately for \jpsiee\ and
\jpsimm.  As for the other final states, the distributions are fit to
obtain the number of mesons.  The resolution smearing parameters are
not required 
to be the same for the two plots, but are consistent:
$1.5\pm0.8$\mevcc\ (\epem) and $1.8\pm0.5$\mevcc\ (\mumu). 
The secondary branching fractions in Eq.~\ref{eq:bfform} are in
this case ${\cal B}_c =
\BR(\psitwos\to\jpsi\pipi)\cdot\BR(\jpsi\to\ellell)$.

\section{Direct Branching Fractions}

To obtain the branching fraction for \jpsi\ mesons produced directly
in the  decay of $B$ mesons, we subtract the feeddown contributions to
the inclusive branching fraction due to the decay of \chicone,
\chictwo, and \psitwos\ mesons.  For the \chicone\ and \chictwo, the
feeddown 
branching fraction is $\BR(B\to\chicj X)\cdot 
\BR(\chicj\to\jpsi\gamma)$, while for the \psitwos, it is 
$\BR(B\to\psitwos X)\cdot \BR(\psitwos\to\jpsi X)$.

Similarly, the feeddown from the \psitwos\ to the \chicone\ and
\chictwo\ is $\BR(B\to\psitwos X)\cdot
\BR(\psitwos\to\chicj\gamma)$. 

Note that a number of uncertainties are
common to both the inclusive and feeddown components, including track
quality and particle identification criteria,
$\BR(\chicone\to\jpsi\gamma)$, and $\BR(\chictwo\to\jpsi\gamma)$.
We use world
average values \cite{ref:pdg2000} for the \psitwos\ branching fractions.
The resulting direct branching fractions are summarized in
Table~\ref{tab:sum}. 

\begin{table}
\caption{Summary of $B$ branching fractions (percent) to charmonium mesons
with statistical and systematic uncertainties. 
The direct branching fraction is also listed, where appropriate.
Last column contains the world average values \cite{ref:pdg2000}.  
}   
\begin{center}
\begin{tabular}{|lcccc|}
\hline \hline
\em  Meson & \em Value &\em Stat &\em Sys & \em World Average \\
\hline 
\jpsi            &  1.057 & $\pm0.012$ & $\pm0.040$ & $1.15 \pm 0.06 $ \\  
\jpsi\ direct    &  0.740 & $\pm0.023$ & $\pm0.043$ & $0.80 \pm 0.08 $\\ 
\chicone         &  0.367 & $\pm0.035$ & $\pm0.044$ & $0.36 \pm 0.05 $\\ 
\chicone\ direct &  0.341 & $\pm0.035$ & $\pm0.042$ & $0.33 \pm 0.05 $\\ 
\chictwo         &  0.210 & $\pm0.045$ & $\pm0.031$ & $0.07 \pm 0.04 $ \\
\chictwo\ direct &  0.190 & $\pm0.045$ & $\pm0.029$ &  -- \\
\psitwos         &  0.297 & $\pm0.020$ & $\pm0.020$ & $0.35 \pm 0.05 $\\ 
\hline \hline
\end{tabular}
\end{center}
\label{tab:sum}
\end{table}

\section{\pstar\ Distributions}

The momentum distributions of charmonium mesons provide an insight
into their production mechanisms.  Since we do not fully reconstruct
the $B$ meson, we cannot determine the meson momentum in the $B$ rest
frame and instead use \pstar, the value in the \FourS\ center-of-mass
frame. The difference, due to the motion of the $B$ in the
center-of-mass frame, has an rms spread of 0.12\gevc.

\subsection{Inclusive \pstar\ Distributions}\label{sub:pstar}

To measure the \pstar\ distributions of \jpsi, \chicone, \chictwo, and
\psitwos\ mesons produced in $B$ decays, we create mass or
mass-difference histograms of on-resonance candidates with \pstar\ in
the desired range.  The \epem\ and \mumu\ final states are again treated 
separately.  The distributions are then fit, with all signal
pdf parameters (other than the number of mesons) fixed to the values
obtained from the earlier fits. 
The fits are performed for 100\mevc\-wide \pstar\ ranges, 
and in each case the sum of
the yields differs from the original fit by fewer than ten events.

In the case of the \jpsi, we perform similar fits on the off-resonance
data and perform a continuum subtraction for each \pstar\ bin.  Since
there are no statistically significant off-resonance \chicone,
\chictwo, or \psitwos\ signals, we do not perform a continuum
subtraction in these cases.

\begin{figure}
\includegraphics[width=\linewidth]{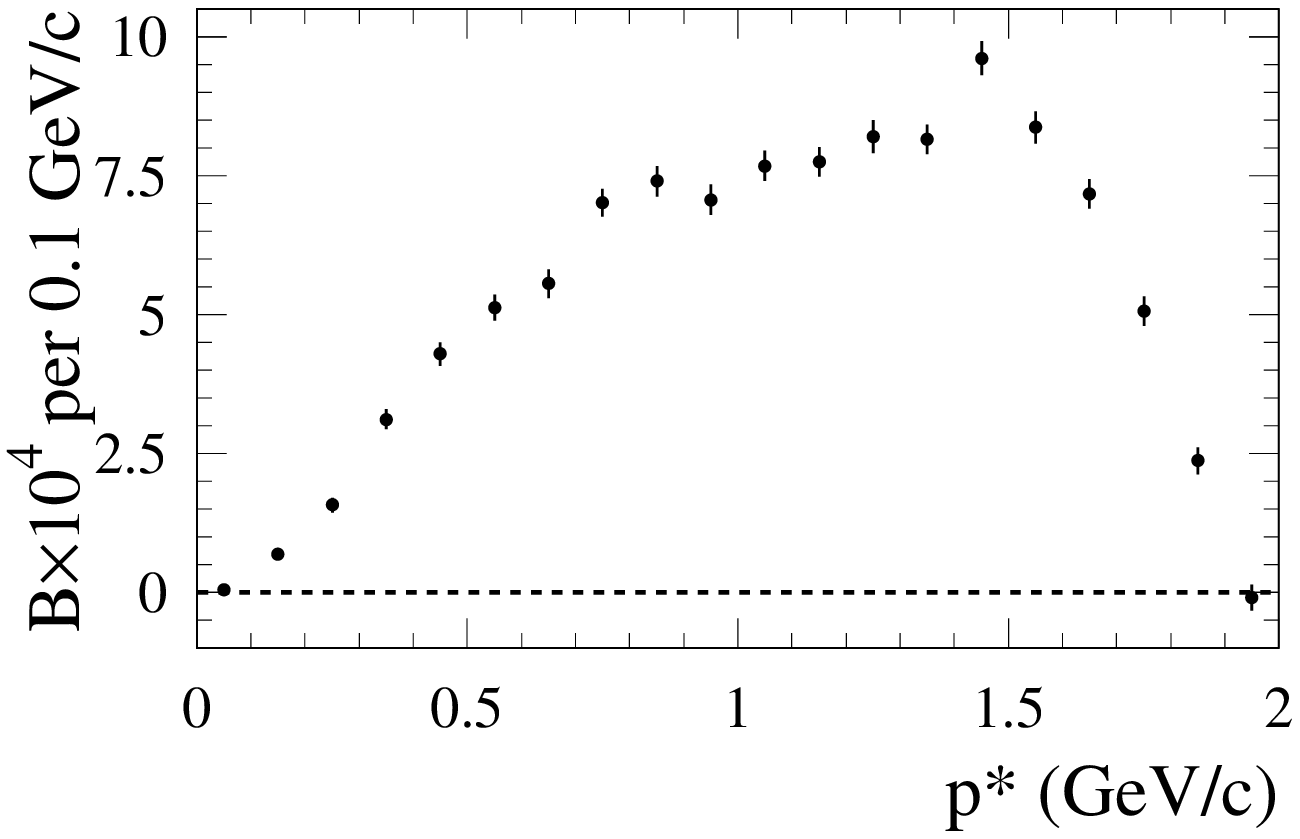}
\caption{$B\to\jpsi X$ branching fraction as a function of \pstar.}
\label{fig:jpsipstar}
\end{figure}

\begin{figure}
\includegraphics[width=\linewidth]{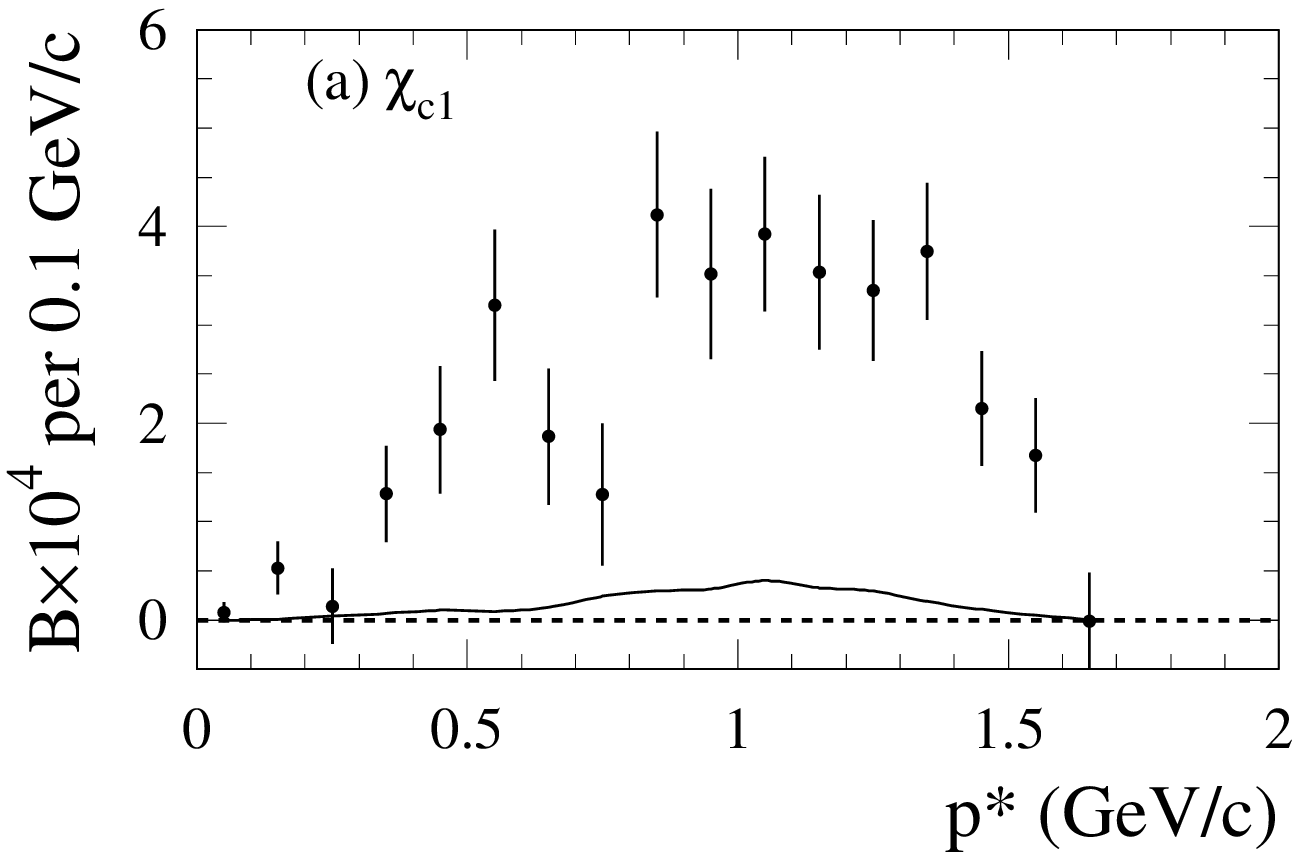}
\includegraphics[width=\linewidth]{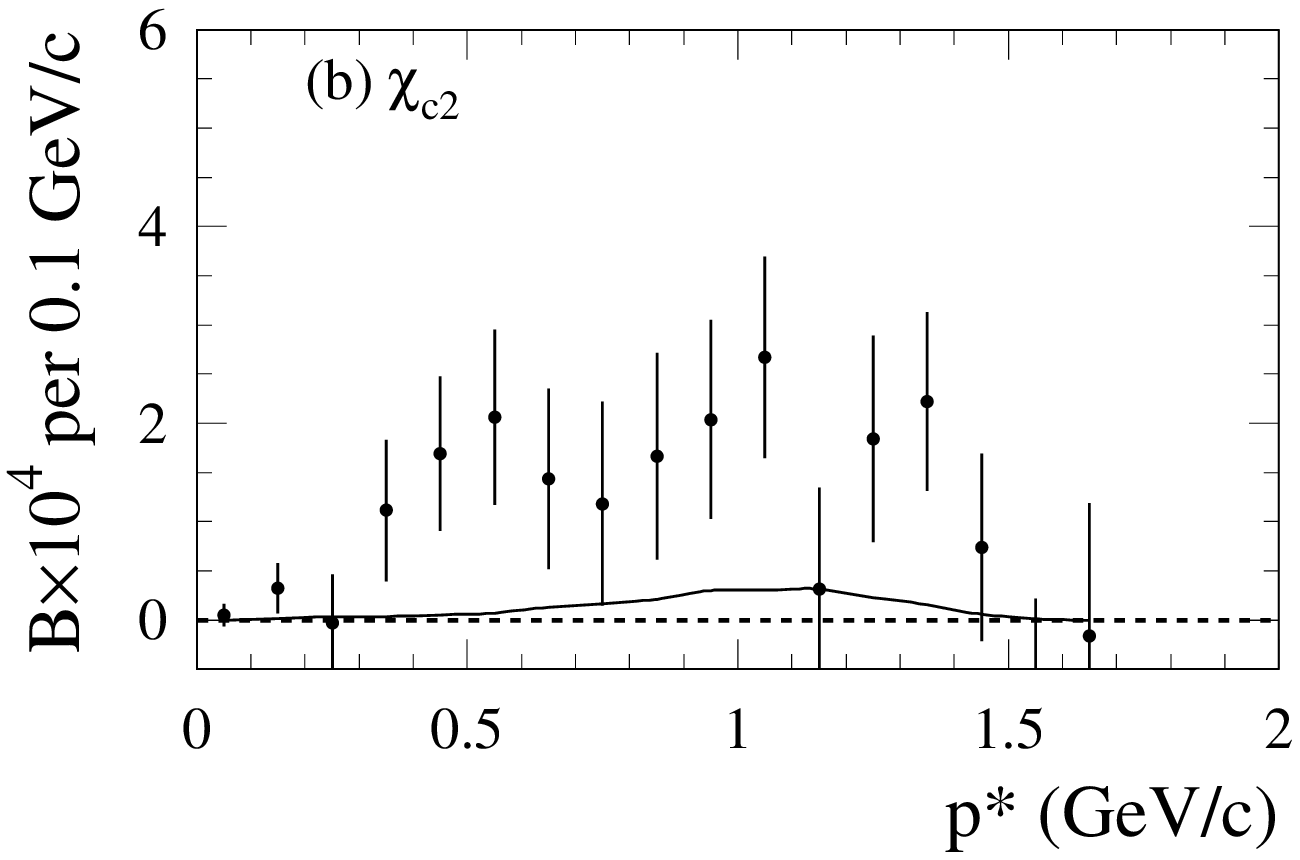}
\caption{Branching fraction as a function of \pstar\ for
(a) $B\to\chicone X$ and (b) $B\to\chictwo X$.
The distribution includes a small feeddown component from
the \psitwos\ (solid curve).}
\label{fig:pstarchic}
\end{figure}

\begin{figure}
\includegraphics[width=\linewidth]{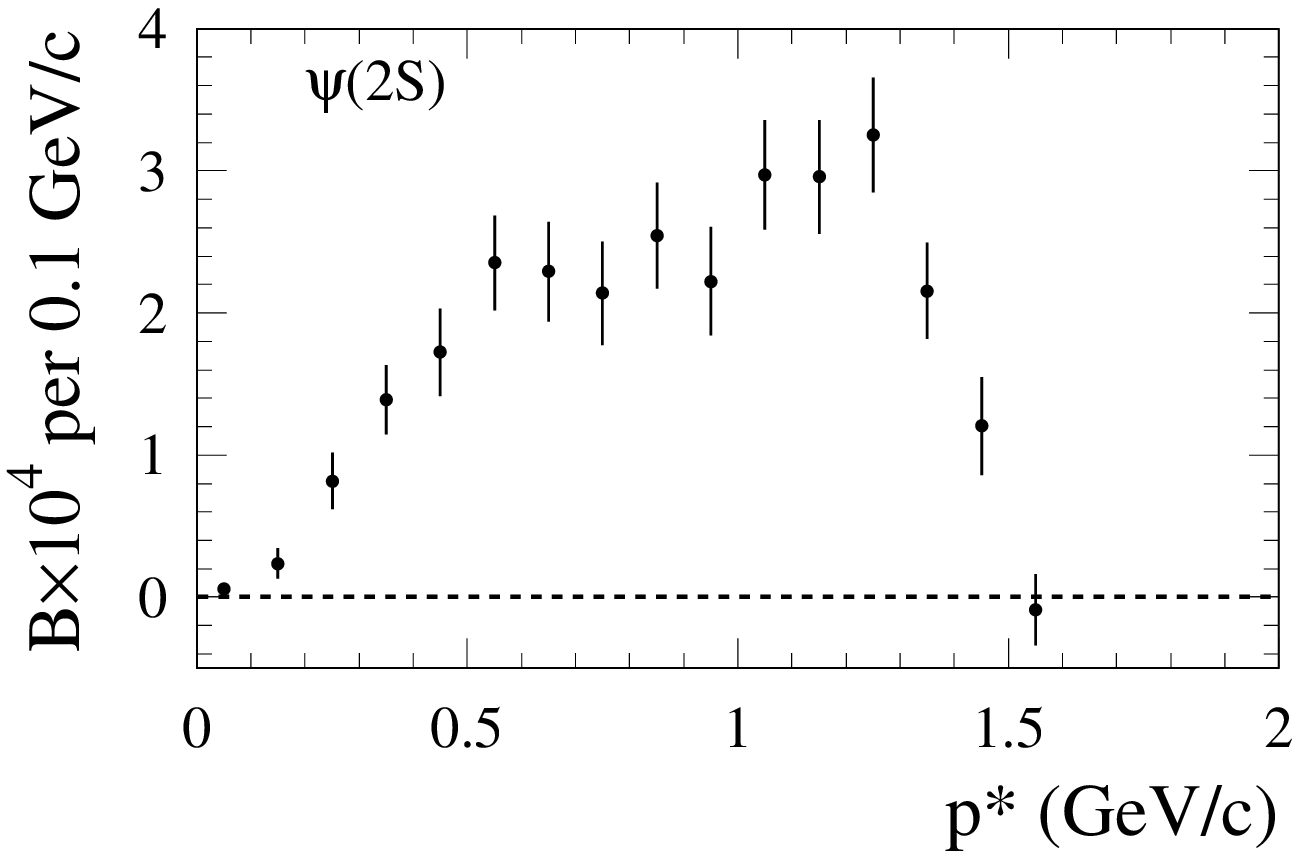}
\caption{$B\to\psitwos X$ branching fraction as a function of \pstar.} 
\label{fig:pstarpsi2s}
\end{figure}

The yield in each bin is corrected by 
the reconstruction efficiency obtained from 
simulated data, which decreases by approximately 10\% between 0 and
2\gevc. The yield is then multiplied by an overall normalization factor
for that particular final state and mode, which
adjusts the
sum of all bins to the earlier branching fraction measurement.
We then perform a weighted average of the two distributions
for the \jpsi, \chicone, or \chictwo, 
or the four distributions for the \psitwos, to
obtain the distributions shown in 
Fig.~\ref{fig:jpsipstar}--\ref{fig:pstarpsi2s}.  For this  purpose, we
use the $\psitwos\to\epem$ and $\psitwos\to\mumu$ branching fractions
from Ref.~\cite{ref:psi2s}.
In all cases, the
distributions that are combined are consistent within statistical errors.  

\subsection{Direct \pstar\ Distributions}

\begin{figure}
\includegraphics[width=\linewidth]{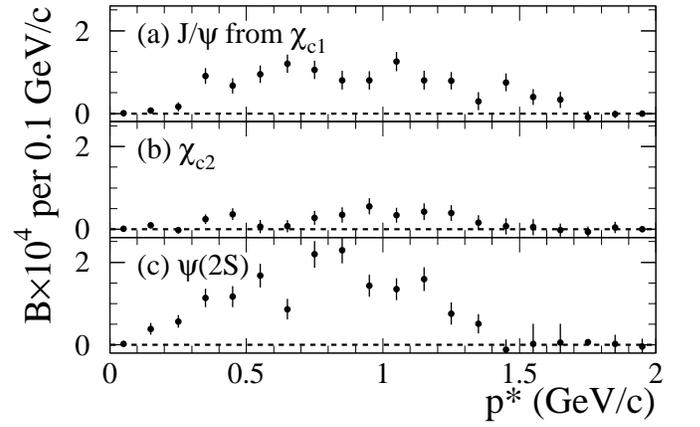}
\caption{Contributions to the $B\to\jpsi X$ branching fraction 
as a function of \pstar\ due to feeddown from (a) \chicone,
(b) \chictwo\, and (c) \psitwos\ mesons.}
\label{fig:jpsifeed}
\end{figure}

The \jpsi\ \pstar\ distribution (Fig.~\ref{fig:jpsipstar}) 
includes components due to 
mesons from the decays 
$\chicone\to\jpsi\gamma$, $\chictwo\to\jpsi\gamma$, and 
$\psitwos\to\jpsi X$.  To measure these distributions, we repeat the
analysis with the data binned by the \pstar\ of the \jpsi\
daughter. The resulting $\jpsi$ feeddown distributions are presented
in Fig.~\ref{fig:jpsifeed}.  

Note that we are using only the $\jpsi\pipi$ decay mode to obtain the
\jpsi\ distribution from \psitwos\ decay.  
In fact, 10.5\% of $\psitwos\to\jpsi
X$ decays are modes other than $\jpsi\pi\pi$.  If we instead use the
simulated \jpsi\ distribution for this 10.5\%,
Fig.~\ref{fig:jpsifeed}c changes by no more than a small fraction of
the statistical error bar in any bin.

\begin{figure}
\includegraphics[width=\linewidth]{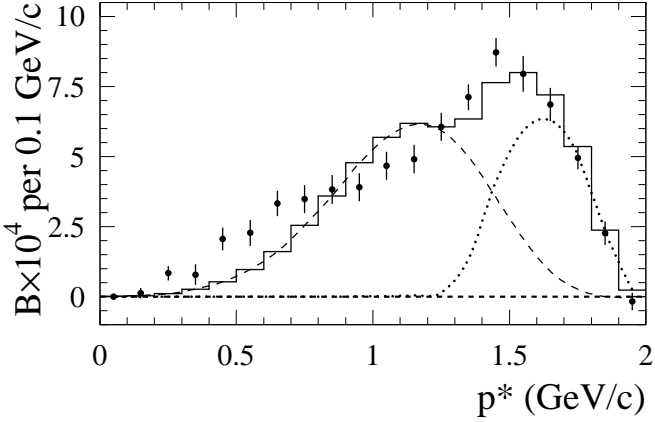}
\caption{\pstar\ of \jpsi\ mesons produced directly in $B$ decays
(points). The histogram is the sum of
the color-octet component from a recent NRQCD 
calculation \cite{ref:benekep} (dashed
line) and the color-singlet $\jpsi K^{(*)}$ component from 
simulation (dotted line).}
\label{fig:directjpsi}
\end{figure}

Subtracting these three components from the inclusive \jpsi\
distribution in Fig.~\ref{fig:jpsipstar} leaves the contribution due
to the \jpsi\ mesons produced directly in $B$ decay
(Fig.~\ref{fig:directjpsi}). 

The superimposed histogram is a calculation of the expected
distribution, which includes color octet and color singlet
components.  We use a recent NRQCD calculation \cite{ref:benekep}
for the color
octet component.  The authors attribute the singlet component to
$\jpsi K^{(*)}$ production, which we obtain from simulation.  The two
are normalized to obtain the best fit to our data.  
Possible sources of the apparent excess at low
momentum are an intrinsic charm component of the $B$ \cite{ref:hou}, 
the production, together with the \jpsi, 
of baryons \cite{ref:brodsky}, 
or an $s{\overline d}g$ hybrid \cite{ref:eilam}.

The small feeddown contribution to \chicone\ and \chictwo\ from
\psitwos\ decay is calculated by simulation and is shown in
Fig.~\ref{fig:pstarchic}. 

\section{\jpsi\ Helicity}

The helicity $\theta_H$ of a $\jpsi\to\ellell$ candidate is the angle,
measured 
in the \jpsi\ rest frame, between the positively charged lepton and
the flight direction of the \jpsi\ in the \FourS\ 
center-of-mass frame.  
A more natural definition would use the $B$ rest frame, but it 
cannot be determined in this analysis.  Simulation indicates that the
rms spread of the difference between the two definitions is
0.085 in \coshel.

\begin{figure}
\includegraphics[width=\linewidth]{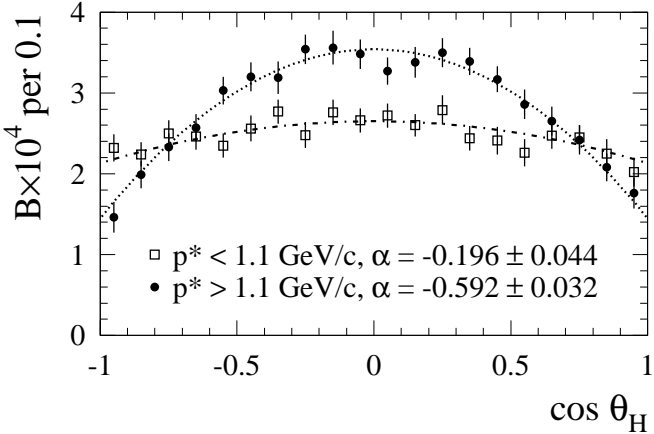}
\caption{Helicity of \jpsi\ mesons produced in $B$ decay with
$\pstar>1.1$~\gevc\ (dots) and $\pstar<1.1$~\gevc\ (open squares).}
\label{fig:jpsihel}
\end{figure}

\subsection{Inclusive Helicity Distribution}

We proceed as for the \jpsi\ \pstar\ distribution, with data
categorized into ranges of width 0.1 in \coshel\ for two different
momentum ranges, which we choose as   
$\pstar<1.1$\gevc\ and $1.1<\pstar<2.0$\gevc.  We fit
the on- and off-resonance 
mass distributions to obtain yields in each bin and
perform a continuum subtraction.  We
correct using the reconstruction
efficiency obtained from simulation for that range, 
although we
observe little dependence of efficiency on helicity. We then 
apply separate normalization factors to the \epem\ and \mumu\ data
such that the total branching fraction (summed over the two \pstar\
ranges) agrees with the value obtained earlier for that mode. 
The distributions from \epem\ and \mumu\
are consistent and are averaged to obtain the helicity distributions
for each of the two \pstar\ ranges (Fig.~\ref{fig:jpsihel}).

We fit each distribution with a function $1+\alpha\cdot\cos^2\theta_H$
to obtain the polarization $\alpha$, where $\alpha=0$ indicates the
sample is unpolarized, $\alpha=1$ transversely polarized, and
$\alpha=-1$ longitudinally polarized.  The high \pstar\ region,
which includes the two-body $B$ decays, is more highly polarized,
$\alpha=-0.592\pm0.032$, than the lower \pstar\ region, $\alpha =
-0.196\pm0.044$. 

We assign a systematic error of 0.008 to these polarizations by
instead considering the reconstruction efficiency to be independent of
helicity. 

\begin{figure}
\includegraphics[width=\linewidth]{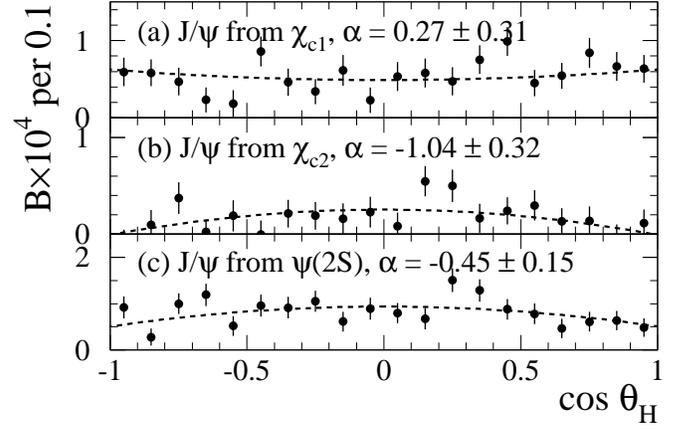}
\caption{Helicity distribution of \jpsi\ mesons produced in the decay
of (a) \chicone, (b) \chictwo, and (c) \psitwos\ mesons.}
\label{fig:helfeed}
\end{figure}

\subsection{Direct \jpsi\ Helicity}

\begin{figure}
\includegraphics[width=\linewidth]{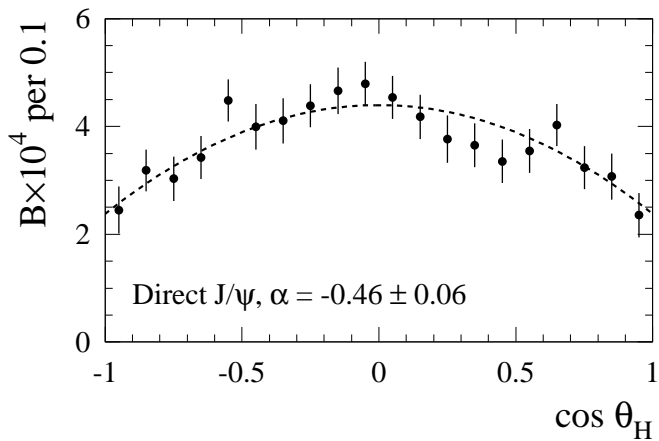}
\caption{Helicity distribution of \jpsi\ mesons produced directly in
the decay of $B$ mesons.}
\label{fig:heldirect}
\end{figure}

We determine the helicity distributions of \jpsi\ mesons produced in
the decay of \chicone, \chictwo, and \psitwos\ in the same way we
calculate 
the \pstar\
feeddown.  Because of the limited statistics of these samples, we
combine the two momentum regions used in the inclusive
analysis.  The resulting feeddown helicity distributions are
shown together with the polarization fits in
Fig.~\ref{fig:helfeed}. We subtract these from the sum of the two
distributions in Fig.~\ref{fig:jpsihel} to obtain the helicity
distribution for the \jpsi\ produced directly in $B$ decay
(Fig.~\ref{fig:heldirect}). The polarization, $\alpha = -0.46\pm0.06$,
is slightly out of the 
range $-0.33$ to 0.05 predicted by an NRQCD 
calculation \cite{ref:fleming}, but other authors have argued
\cite{ref:ma} that relativistic corrections reduce the reliability of
the calculation.  The systematic uncertainty of 0.008 obtained above
is small compared to the statistical error.  This result is difficult
to compare directly with that from CDF
\cite{ref:cdfpol}, due to the different mixture of $b$
mesons and baryons, and the distinction between the effective helicity
calculated there and the true helicity.

\section{Summary}

We have reported new measurements of $B$ meson decays 
to final states including charmonium mesons, which are summarized in
Table~\ref{tab:sum}. 
We 
have presented a number of momentum distributions.  The distributions
of the feeddown \jpsi\  daughters of the \chicone, \chictwo, and
\psitwos\ have not previously been measured, and allow us to more
accurately determine the distribution for \jpsi\ mesons produced
directly in $B$ decay. The  direct \jpsi\ distribution
is compared to a recent
NRQCD calculation and appears to indicate an excess at low momentum.

The \jpsi\ helicity distribution, which has also has not previously
been published, indicates that the polarization of direct \jpsi\
mesons is slightly out of the range predicted by an NRQCD
calculation.

We are grateful for the excellent luminosity and machine conditions
provided by our \pep2\ colleagues, 
and for the substantial dedicated effort from
the computing organizations that support \babar.
The collaborating institutions wish to thank 
SLAC for its support and kind hospitality. 
This work is supported by
DOE
and NSF (USA),
NSERC (Canada),
IHEP (China),
CEA and
CNRS-IN2P3
(France),
BMBF and DFG
(Germany),
INFN (Italy),
NFR (Norway),
MIST (Russia), and
PPARC (United Kingdom). 
Individuals have received support from the 
A.~P.~Sloan Foundation, 
Research Corporation,
and Alexander von Humboldt Foundation.


\begin{thebibliography}{99}


\bibitem{ref:bodwin}
G. T. Bodwin, E. Braaten, and G. P. Lepage, \jprd {51}, 1125
(1995); Erratum {\bf 55}, 5853 (1997); M.\ Beneke, F.\ Maltoni, and
I.\ Z.\ Rothstein, \jprd{59}, 054003 (1999). 

\bibitem{ref:psi2sbrat}
E.\ Braaten and S.\ Fleming, \jprl {74}, 3327 (1995); M.~Cacciari,
M.~Greco, M.L.~Mangano, and A.~Petrelli, \plb {356}, 553 (1995).

\bibitem{ref:abe}
CDF Collaboration, F. Abe {\em et al.}, \jprl {69}, 3704 (1992);
{\bf 79}, 572 (1997); {\bf 79}, 578 (1997). D0 Collaboration, S.\
Abachi {\em et al.}, \plb {370}, 239 (1996).

\bibitem{ref:leibovich}
A.\ K.\ Leibovich, \npps {93}, 182 (2001); G.\ A.\ Schuler, 
\epjc{8}, 273 (1999).

\bibitem{ref:cleo}
CLEO Collaboration, R.\ Balest {\em et al.}, \jprd{52}, 2661 (1995); 
S.\ Chen {\em et al.}, \jprd{63}, 031102 (2001).

\bibitem{ref:bellechic}
Belle Collaboration, K.\ Abe {\em et al.}, \jprl{89}, 011803 (2002).

\bibitem{ref:babar}
\babar\ Collaboration, B.\ Aubert {\em et al.},
\nim{A479}, 1 (2002).

\bibitem{ref:geant}
``GEANT Detector Description and Simulation Tool'', Version 3.21, CERN
Program Library Long Writeup W5013 (1994).

\bibitem{ref:fox}
G.~C.~Fox and S.~Wolfram, \jprl{41}, 1581 (1978).

\bibitem{ref:jetset}
T.\ Sjostrand, Computer Physics Commun.\ 82 (1994) 74.

\bibitem{ref:lat}
A.\ Drescher {\em et al.}, \nim{A237}, 464 (1985). ${\rm LAT} = 
{\sum_{i=3}^{n} E_i r_i^2}/ 
{(E_1 r_0^2 + E_2 r_0^2 + \sum_{i=3}^{n}E_i r_i^2)}$, where the $n$
crystals in the EMC cluster are ranked in order of energy deposited in
that crystal, $E_i$,
and $r_0 = 5$\cm\ is the average distance between crystal
centers. $r_i$ is the distance between crystal $i$ and the cluster 
centroid calculated from an energy-weighted average of the $n$
crystals. 

\bibitem{ref:zern}
R.\ Sinkus and T.\ Voss, \nim{A391}, 360 (1997). 
The Zernike moment $A_{nm}$ is calculated using the energy $E_i$ and
location $(\rho_i, \phi_i)$ of crystals with respect to the shower
centroid. The location 
is defined in a cylindrical coordinate system with the $z$ axis
running from 
the beam spot to the centroid, where $\rho_i = r_i/R_0$ and $R_0 =
15$\cm. $A_{nm} = \sum_{i=1}^n (E_i/E)\cdot f_{nm}(\rho_i)
e^{-im\phi_i}$, where the sum includes only crystals with 
$\rho_i\le 1$, and $E$ is the total energy in the cluster. The Zernike
functions are 
$f_{nm}(\rho) = \sum_{s=0}^{(n-m)/2}
\frac{(-1)^s(n-s)!\rho^{n-2s}}
{s!((n+m)/2 - s)!((n-m)/2 - s)!}$, with $m\le n$ and $(n-m)$
even. Studies indicate that $A_{42}$ provides good  
separation between hadronic and electromagnetic showers when used in
conjunction with LAT.

\bibitem{ref:cont}
\babar\ Collaboration, B.\ Aubert {\em et al.}, \jprl{87}, 162002 (2001);
Belle Collaboration, K.\ Abe {\em et al.}, \jprl{88}, 052001 (2002).

\bibitem{ref:photos}
E.~Barberio, B.~van Eijk, and Z.~Was, Comput.\ Phys.\ Commun.\ {\bf
66}, 115 (1991).

\bibitem{ref:pdg2000}
Particle Data Group, K.\ Hagiwara {\em et al.}, \jprd{66}, 010001 (2002).

\bibitem{ref:bodwinc2}
G.T.\ Bodwin, E.\ Braaten, T.C.\ Yuan, and G.P.\ Lepage, 
\jprd{46}, 3703 (1992).

\bibitem{ref:kuhn}
J.\ H.\ K\"{u}hn, S.\ Nussinov, and R.\ R\"{u}ckl, Z.\ Phys.\ {\bf C
5}, 117 (1980). 

\bibitem{ref:psi2s}
\babar\ Collaboration, B.\ Aubert {\em et al.}, \jprd{65}, 031101 (2002).

\bibitem{ref:bes}
BES Collaboration, J.~Z.~Bai {\em et al.}, \jprd{62}, 032002 (2000).

\bibitem{ref:benekep}
M.\ Beneke, G.A.\ Schuler, and S.\ Wolf, \jprd{62}, 034004 (2000).
Curve is from Fig.~5, $\Lambda = 300$\mev, $p_F=300$\gev, and 
$m_{sp}=150$\mev.

\bibitem{ref:hou}
C.-H.V.\ Chang and W.-S.\ Hou, \jprd{64}, 071501 (2001).

\bibitem{ref:brodsky}
S.J.\ Brodsky and F.S.\ Navarra, \plb{411}, 152 (1997).

\bibitem{ref:eilam}
G.\ Eilam, M.\ Ladisa, and Y.-D.\ Yang, \jprd{65}, 037504 (2002).

\bibitem{ref:fleming}
S.\ Fleming {\em et al.}, \jprd{55}, 4098 (1997).

\bibitem{ref:ma}
J.\ P.\ Ma, \jprd{62}, 054012 (2000).

\bibitem{ref:cdfpol}
CDF Collaboration, T.\ Affolder {\em et al.}, \jprl {85}, 2886 (2000).

\end{thebibliography}
\end{document}